\documentclass[english,floatfix,showkeys,superscriptaddress,nofootinbib,aps,prd]{revtex4}
\usepackage[latin1]{inputenc}
\usepackage[T1]{fontenc}
\usepackage{lmodern}
\usepackage{amsmath}
\usepackage{amssymb}
\usepackage{graphicx}
\usepackage{esint}
\usepackage{longtable}
\usepackage{dcolumn}
\usepackage{babel} 
\usepackage{csquotes} 
\usepackage{color} 
\usepackage{setspace}

\begin{document}

\title{Kasner cosmology in bumblebee gravity}

%%%%%%%%%%%%%%%%%%%%%%%%%%%%%%%%%%%%%%%%%%%%%%%%%%%%%%%%%%%%%%%%%%%%%%

\author{Juliano C. S. Neves} 
\email{juliano.c.s.neves@gmail.com}
\affiliation{Instituto de Ci\^encia e Tecnologia, Universidade Federal de Alfenas, \\ Rodovia Jos\'e Aur\'elio Vilela,
11999, CEP 37715-400 Po\c{c}os de Caldas, MG, Brazil}

\begin{abstract}
Kasner cosmology is a vacuum and anisotropically expanding spacetime in the general relativity context. 
In this work, such a cosmological model is studied in another context, the bumblebee model, where the 
Lorentz symmetry is spontaneously broken. By using the bumblebee context it is possible
to justify the anisotropic feature of the Kasner cosmology. Thus, the origin of the anisotropy in this 
 cosmological model could be in the Lorentz symmetry breaking. Lastly, an application in the pre-inflationary
 cosmology is suggested.
\end{abstract}


\maketitle

\section{Introduction}
Kasner geometry was obtained in 1921 \cite{Kasner} from a problem of embedding geometries 
into a flat and higher dimensional spacetime. 
Later it was conceived of as a cosmological model.\footnote{See, for example, Ref. \cite{Gravitation},
 chapter 30, and references therein.}
The weirdness of the Kasner cosmological model is due to the fact that such spacetime is a vacuum and anisotropic
solution of Einstein's field equations that describes an expanding universe. Since then, the Kasner metric
has been adopted in several studies of anisotropic cosmologies as some sort of limiting case \cite{Gravitation}. 
Studies on Kasner geometry have been made even in contexts beyond general relativity, like the $f(T)$ model 
\cite{Paliathanasis:2017htk,Skugoreva:2017vde}, brane-world model \cite{Frolov:2001wz} and
loop quantum gravity \cite{Gupt:2012vi}.

Here the main intention is to show that the Kasner metric is a solution 
of the modified field equations in a different context:
the bumblebee gravity, which is a scenario of Lorentz symmetry breaking.
Several contexts or scenarios deal with the violation of one of the most important 
symmetries in physics, like string theory \cite{Kostelecky:1988zi,Kostelecky:1989jw}, 
 noncommutative spacetime \cite{Carroll:2001ws,Ferrari:2006gs},
brane worlds \cite{Santos:2012if} and Ho\v rava--Lifshitz gravity \cite{Horava:2009uw}.
The initial steps to the Lorentz violation in the gravitational context were given by 
Kosteleck\'y \cite{Kostelecky2004}, who developed the so-called standard model extension, something
initially created to the particles phenomenology \cite{Colladay:1996iz,Colladay:1998fq}. 
In such a context involving gravity, the spontaneous symmetry breaking 
preserves both the geometric constraints and conservation laws or quantities required by the theory of general relativity
or a Riemannian geometry.
Another alternative to the Lorentz violation, even using the standard model extension, is to adopt
the Finsler geometry, in which the symmetry violation is obtained directly from geometric structures of 
spacetime.\footnote{See, for example, Ref. \cite{Silva:2019qzl}.}  

Recently, several articles have described solutions of the modified gravitational field equations in the bumblebee
scenario. Black hole and wormhole solutions \cite{Casana,Maluf_Neves,Bertolami,Maluf:2021ywn,Oliveira:2021abg,Gullu:2020qzu,Ovgun:2018xys,Lessa:2019bgi,Lessa:2020imi}, cosmological models \cite{Capelo,Petrov,Jesus,ONeal,MalufNeves,MalufNeves2}, and  
gravitational waves \cite{Kostelecky:2016kkn,Kostelecky:2016kfm}  have shown the influence of the
Lorentz-violating parameter on the geometry or even on phenomena like the cosmic acceleration \cite{Capelo,MalufNeves,MalufNeves2} and the black hole shadow \cite{Maluf_Neves}. 
In Refs. \cite{Casana,MalufNeves,Paramos:2014mda,Gu:2022grg} bounds on that parameter are 
calculated from the gravitational
sector by means of several approaches.\footnote{However, Maluf and Muniz \cite{Maluf:2022knd} 
have pointed out that the Kerr-like geometry adopted in Ref. \cite{Gu:2022grg} is not a correct solution.} 
But as for the subject
of this article, the main focus is on the anisotropic cosmic expansion in which each spatial direction expands
differently when each one is compared to others.  

As I said, models of Lorentz symmetry breaking adopt modified field equations for
the gravitational sector. As we will see, Kasner cosmology could be solution of those modified equations.
Also the Kasner parameters or exponents could contain the Lorentz-violation parameter, thus
it is argued that \textit{the origin of the Kasner geometry or its anisotropic 
feature is due to the Lorentz symmetry breaking}.
Apart from \enquote{realistic} models in bumblebee gravity studied in the mentioned references, 
like our work \cite{MalufNeves}, 
here the main idea is providing a theoretical justification for the Kasner cosmology and suggesting that a
highly anisotropic spacetime like Kasner's could be useful in a pre-inflationary discussion, where symmetries
could be broken.
As we will see, each spatial direction expands or 
contracts differently, and they are related to the Lorentz-violating parameter.

The article is structured as follows: Sec. 2 presents briefly the model adopted here, and in Sec. 3
the Kasner metric in the bumblebee scenario is obtained. The final comments are in Sec. 4.
In this article, $G=c=1$, where $G$ is the gravitational constant, and $c$ is speed of light in vacuum.
Greek index runs from 0 to 3, and Latin index runs from 1 to 3.

\section{The modified field equations}
 The action for the bumblebee model in the simplest way---not including torsion and including a vector field
 coupled to the geometry---is given by
\begin{equation}
 S_{B}=\int d^{4}x\sqrt{-g} \bigg [ \frac{R}{2\kappa}+\frac{\xi}{2\kappa}B^{\mu}B^{\nu}R_{\mu\nu}-\frac{1}{4} B_{\mu\nu}B^{\mu\nu} -V(B^{\mu}B_{\mu}\pm b^{2})+\mathcal{L}_{M} \bigg ],
 \label{S1} 
\end{equation}
where $\sqrt{-g}$ is the metric determinant, $\kappa=8\pi G/c^4$, $\xi$ is the coupling constant, 
$R$ is the Ricci scalar, $R_{\mu\nu}$ is the Ricci tensor, $B_\mu$ is the 
 bumblebee vector field, $V$ is the field potential, 
and $\mathcal{L}_M$ stands for the matter Lagrangian. 
The bumblebee potential has an important role in this problem, it is responsible for triggering a nonzero 
vacuum expectation value (VEV) for the vector field. That is to say, 
as the bumblebee assumes a nonzero VEV $\langle B_{\mu}\rangle=b_\mu\neq 0$, the Lorentz
symmetry is spontaneously broken. In particular, in this vacuum state $b_{\mu}b^{\mu}=\pm b^2$, thus  $V=0$.
According to the metric signature adopted here, the $\pm$ values indicated in the potential mean either timelike or spacelike values, respectively, 
 for the norm of $b_{\mu}$.
 Another important element in the action (\ref{S1}) is the field 
 strength 
\begin{equation}
B_{\mu\nu}=\partial_{\mu}B_{\nu}-\partial_{\nu}B_{\mu},
\end{equation}
 which will be zero because of the type of field chosen here. 

The gravitational field equations for the bumblebee model are calculated by varying the action (\ref{S1}) with respect
to the metric field $g_{\mu\nu}$. This valuable procedure delivers the following equations:
\begin{align}
G_{\mu\nu}  = & \ R_{\mu\nu}-\frac{1}{2}g_{\mu\nu}R = \kappa \left( T^{B}_{\mu\nu}+ T^{M}_{\mu\nu} \right) \nonumber \\
  = & \ \kappa\left[2V'B_{\mu}B_{\nu} +B_{\mu}^{\ \alpha}B_{\nu\alpha}-\left(V+ \frac{1}{4}B_{\alpha\beta}B^{\alpha\beta}\right)g_{\mu\nu} \right]  +  \xi \bigg[ \frac{1}{2}B^{\alpha}B^{\beta}R_{\alpha\beta}g_{\mu\nu}-B_{\mu}B^{\alpha}R_{\alpha\nu}  \nonumber\\
  & -B_{\nu}B^{\alpha}R_{\alpha\mu}  + \frac{1}{2}\nabla_{\alpha}\nabla_{\mu}\left(B^{\alpha}B_{\nu}\right)+\frac{1}{2}\nabla_{\alpha}\nabla_{\nu}\left(B^{\alpha}B_{\mu}\right)   -  \frac{1}{2}\nabla^{2}\left(B_{\mu}B_{\nu}\right)-\frac{1}{2}
g_{\mu\nu}\nabla_{\alpha}\nabla_{\beta}\left(B^{\alpha}B^{\beta}\right) \bigg] \nonumber\\
 &  +\kappa T^{M}_{\mu\nu},
\label{modified}
\end{align}
where $G_{\mu\nu}$ is the Einstein tensor, and the operator $'$ means derivative with
respect to the potential argument (according to the vacuum condition assumed here, $V=V'=0$). 
Lastly, the energy-momentum tensor of the matter fields is $T_{\mu\nu}^{M}$, and the corresponding tensor for
the bumblebee vector field is $T_{\mu\nu}^{B}$. Then the main procedure here is solving the modified
field equations (\ref{modified}) for a specific vector field $B_{\mu}$, from a specific geometry. 
 
 By varying the action (\ref{S1}) with respect to the bumblebee field, 
 such a procedure provides an equation of motion for the vector field $B_{\mu}$, namely
 \begin{equation}
 \nabla_{\mu}B^{\mu\nu}=2\left( V'B^{\nu}-\frac{\xi}{2\kappa}B_\mu R^{\mu\nu}  \right).
 \label{B_eq}
 \end{equation}
Any spacetime or metric that is solution of the modified field equations (\ref{modified})
also must be solution of the vector field equations (\ref{B_eq}).

\section{Solving the modified field equations}
 
 \subsection{Kasner cosmology} 
The Kasner  metric \cite{Kasner} is given by the following line element:
\begin{equation}
ds^2=-dt^2+t^{2p_1}\left(dx^1\right)^2+t^{2p_2}\left(dx^2\right)^2+t^{2p_3}\left(dx^3\right)^2,
\label{Metric}
\end{equation}
 where the exponents $p_1,p_2$ and $p_3$ are constant parameters satisfying two important relations:
 \begin{align}
 p_1+p_2+p_3& = 1,  \label{Condition1} \\
 \left(p_1 \right)^2+\left(p_2 \right)^2+\left(p_3 \right)^2 & = 1.
 \label{Condition2}
\end{align}   
The relation (\ref{Condition1}) is defined in order to provide a
three dimensional flat spacetime for $t$ constant. 
Possible values for the three Kasner exponents, 
in agreement with the relation  (\ref{Condition1}) and even with (\ref{Condition2}), are
\begin{align}
p_2 = & \ \frac{1}{2}\left(1-p_1 \mp \sqrt{1+\left(2-3p_1 \right)p_1} \right),\label{p2} \\
p_3= & \ \frac{1}{2}\left(1-p_1 \pm \sqrt{1+\left(2-3p_1 \right)p_1} \right).\label{p3}
\end{align}  
It is worth pointing out that Eq. (\ref{Metric}) is a vacuum solution in the general relativity context, i.e.
$T_{\mu\nu}^M=0$, and describes an expanding world with its respective volume element given by
\begin{equation}
\sqrt{-g}=t.
\end{equation} 
As time passes, the volume increases anisotropically, for each direction expands at different rates. There are
two expanding directions and one contracting direction.
For sure, Kasner geometry is a weird solution of the Einstein field equations. How can an expanding universe be
anisotropic and empty? So here a justification for that weirdness is proposed.
 As we will see, the bumblebee field will be source for each exponent
$p$ in the Kasner metric.

\subsection{The bumblebee field as source of anisotropies}

For the purpose mentioned above, one assumes that both  $V=V'=0$ and that the VEV of the bumblebee field is given by
\begin{equation}
B_{\mu}=b_{\mu}= \left(\sqrt{\frac{p_i}{\xi}},0,0,0\right),
\label{Field}
\end{equation} 
for $i=1,2$ or 3, where $p_1,p_2$, and $p_3$ are the exponents of the Kasner metric. 
Thus, the field $b_{\mu}$ is a timelike vector field, and the field strength is $B_{\mu\nu}=b_{\mu\nu}=0$.
 Assuming then that the VEV of the bumblebee field is constant and that $b_{\mu}b^{\mu}=-b^2$ for a timelike
 vector, one has
\begin{equation}
b_{\mu}b^{\mu}=-b^2=-\frac{p_i}{\xi}.
\label{Norm}
\end{equation}  
The norm of the bumblebee field or its VEV could depend on any value of $i$, for $i=1,2$ or 3. As we will
see, once one chooses the value of $i$, $p_i$ will be related to a special parameter, the Lorentz-violating parameter.
Indeed, the above result will be useful later because the exponent $p_i$ will be conceived of as the Lorentz-violating
parameter.

Following Casana \textit{et al.} \cite{Casana}, the main idea here is looking for the vacuum version of the modified
field equations (\ref{modified}). In order to obtain that version,  firstly one calculates the trace of the
gravitational field equations. Thus, 
\begin{equation}
R=\xi \nabla_{\alpha} \nabla_{\beta} \left(b^{\alpha}b^{\beta} \right),
\label{Trace}
\end{equation}
where both the condition for the bumblebee field potential and $T^{M}_{\mu\nu}=0$ were assumed. Then by
substituting the trace (\ref{Trace}) into Eq. (\ref{modified}),  the trace-reversed field equations are revealed, namely
\begin{equation}
R_{\mu\nu}=\kappa T^{B}_{\mu\nu}+\frac{\xi}{2}g_{\mu\nu} \nabla_{\alpha}\nabla_{\beta} \left(b^{\alpha} b^{\beta} \right).
\label{Ricci}
\end{equation} 
The vacuum version of the modified field equations is then assumed, that is to say,
\begin{align}
\bar{R}_{\mu\nu}=&  \ R_{\mu\nu} +\xi \bigg[ b_{\mu}b^{\alpha}R_{\alpha\nu} +b_{\nu}b^{\alpha}R_{\alpha\mu} -\frac{1}{2}b^{\alpha}b^{\beta}R_{\alpha\beta}g_{\mu\nu}  -\frac{1}{2}\nabla_{\alpha} \nabla_{\mu}\left(b^{\alpha}b_{\nu} \right)- \frac{1}{2}\nabla_{\alpha}\nabla_{\nu}\left(b^{\alpha}b_{\mu} \right)  \nonumber \\
&  + \frac{1}{2}\nabla^2 \left(b_{\mu}b_{\nu} \right)\bigg] =  0.
\label{Vacuum_equations}
\end{align}
The mathematical trick that provides vacuum field equations considers then the bumblebee influence on the left
side of the field equations, like a geometric influence.  
 
The \textit{Ansatz} (\ref{Metric}) adopted in the vacuum equations (\ref{Vacuum_equations}) and, at the same time, 
the bumblebee field (\ref{Field}) give us four equations written as
\begin{align}
\bar{R}_{00}= & -\frac{1}{2t^2} \left(p_i^2+p_j^2+p_k^2-p_i-p_j-p_k \right)\left(2-3p_i \right), \label{R00} \\
\bar{R}_{ii} = & -\frac{p_i}{ 2 t^{2(1- p_i)}}\bigg[ p_i^2-p_j^2-p_k^2 +2\left(p_j+p_k -\frac{3}{2}\right)p_i  -p_j-p_k+2 \bigg], \label{R11} \\
\bar{R}_{jj}= & \  \frac{1}{ 2 t^{2(1- p_j)}}\bigg[p_i^3- \left(1+2p_j \right)p_i^2- \bigg(p_j^2-p_k^2-3p_j +p_k+2p_jp_k \bigg)p_i  -2 \left(1-p_j-p_k \right)p_j \bigg], \label{R22} \\
\bar{R}_{kk}= &\  \frac{1}{ 2 t^{2(1- p_k)}} \bigg[p_i^3- \left(1+2p_k \right)p_i^2- \bigg(p_k^2-p_j^2-3p_k  +p_j+2p_jp_k \bigg)p_i -2 \left(1-p_j-p_k \right)p_k \bigg], \label{R33}
\end{align}
where $i,j,k=1,2$ or $3$ with $i\neq j\neq k$, and the index $i$ is the same of the bumblebee field given by
 Eq. (\ref{Field}).
The system presented by Eqs.(\ref{R00})-(\ref{R33}) is an overdetermined system. But interestingly 
the relations (\ref{p2}) and (\ref{p3}) are even solutions of the vacuum equations (\ref{R00})-(\ref{R33}). 
Therefore, by using the norm of the bumblebee field, given by Eq. (\ref{Norm}), one has
\begin{align}
p_i=& \ \ell, \label{pi} \\
p_{j,k} = & \ \frac{1}{2}\left(1-\ell \pm \sqrt{1+\left(2-3\ell \right)\ell} \right), \label{pjpk}
\end{align}
where the  parameter $\ell$ is the so-called Lorentz-violating parameter, 
which is commonly defined as $\ell=\xi b^2$, whose best upper bound is $\ell < 10^{-23}$ to date
 \cite{Paramos:2014mda}. 
Thus, the exponents of the Kasner metric are dimensionless. 
Therefore, the vacuum equations give us the parameters for the Kasner cosmology in the bumblebee scenario,
whose real values are obtained for $-1/3<\ell <1$, and, according to which, 
the conditions (\ref{Condition1})-(\ref{Condition2}) are straightforwardly satisfied.
As we can see, now all parameters of the Kasner metric ($p_i,p_j,p_k$) carry the Lorentz-violating parameter
$\ell$. For $\ell=0$ or $b^2=0$, one has $p_i=p_k=0$ and $p_j=1$, and, from the coordinate transformations
$t'=t\cosh x^j$ and $x'^{j}=t \sinh x^j$, the Minkowski metric is again restored like in the general relativity context.

It is worth pointing out that the procedure presented here, in which the parameters (\ref{pi}) and (\ref{pjpk}) 
are solutions of the vacuum equations (\ref{Vacuum_equations}), 
works just for a timelike bumblebee field, like the one indicated in Eq. (\ref{Field}). 
As we will see, a timelike bumblebee field could generate  three different Hubble parameters. In 
Ref. \cite{MalufNeves}, where the Bianchi I cosmology was studied in the Lorentz symmetry breaking context,
just one Hubble parameter is different from the other two because the bumblebee field, which is spacelike in the
mentioned reference, 
points toward one specific spatial direction. Another difference from Ref. \cite{MalufNeves} regards
the matter content. The mentioned article focuses on a universe made up of matter
and radiation that evolves into a matter-dominated universe, something appropriate to the late cosmology. 
On the other hand, as will see, this work focuses on the initial period of the universe, 
something even before the inflationary period. That is the reason why the Kasner spacetime, or 
a vacuum cosmological model, is more suitable to the very early cosmology.   

As I said, the Kasner cosmology in the bumblebee scenario is a vacuum solution, but here it is a vacuum solution
of the modified field equations (\ref{Vacuum_equations}). 
Even the bumblebee field equations (\ref{B_eq}) are satisfied from the metric (\ref{Metric}) and
the parameters  (\ref{pi})-(\ref{pjpk}). Every component of the bumblebee equation (\ref{B_eq}) is identically zero
regardless the mentioned parameters, except the $0$-component, given by 
$\left(p_i^2+p_j^2+p_k^2-p_i-p_j-p_k \right)\sqrt{\xi p_i}/\kappa t^2$, which is also zero for those specific parameters. 

The Kasner metric is a particular case of the Bianchi I geometry, which is written as
\begin{equation}
ds^2=-dt^2+a_{1}(t)^2 \left(dx^1 \right)^2+a_{2}(t)^2 \left(dx^2 \right)^2+a_{3}(t)^2 \left(dx^3 \right)^2,
\label{Bianchi}
\end{equation}
where each spatial direction has its own scale factor $a(t)$. By comparing Eq. (\ref{Metric}) with Eq. (\ref{Bianchi}),
and assuming the exponents (\ref{pi}) and (\ref{pjpk}), one has
\begin{align}
a_i(t) & =t^{\ell}, \label{ai} \\
a_{j,k} (t) & = \left(t^{1-\ell \pm \sqrt{1+\left(2-3\ell \right)\ell}}\right)^{\frac{1}{2}}.\label{ajak}
\end{align}  
Again, $\sqrt{-g}=t$ for the metric (\ref{Bianchi}) with the scale factors (\ref{ai}) and (\ref{ajak}).
And the corresponding directional Hubble parameters are
\begin{align}
H_i (t) & = \frac{\dot{a}_i(t)}{a_i(t)}=\frac{\ell}{t}, \label{Hi} \\
H_{j,k} (t) & = \frac{\dot{a}_{j,k}(t)}{a_{j,k}(t)}= \frac{1- \ell \pm \sqrt{1+\left(2-3\ell \right)\ell}}{2t},\label{Hjk}
\end{align}
where the dot operator  means derivative with respect to time.
As we can see from (\ref{Hi}) and (\ref{Hjk}), assuming $0< \ell \ll 1$, 
there are two expanding directions, $H_i(t)$ and $H_j(t)$, 
and the $k$-direction shrinks ($H_k(t)<0$),
like the Kasner cosmology in the general relativity context. The Lorentz-violating parameter $\ell$ increases
the Hubble parameter for one expanding direction and decreases it for another expanding direction. 
On the other hand, $\ell$ decreases the value of
the Hubble parameter for the contracting direction.
But it is worth emphasizing that each direction expands differently, even choosing a timelike bumblebee field, something contrary to our work  \cite{MalufNeves}, where the directions
in which the bumblebee vector field is absent expand likewise. But the mentioned article speaks of 
a \enquote{realistic} model in Bianchi I cosmology, that is, there are matter fields in that work, something
absent here.

The last point to be addressed is the isotropization of the Kasner (or Bianchi I cosmology) in bumblebee gravity. 
As we could expect, the cosmology presented here is not able to pass through an isotropization
process. In order to see that, anisotropies should be quantified. 
According to Refs. \cite{Bronnikov:2004nu,Saha:2006iu}, the process of becoming isotropic is measured according
to the following criterion:
\begin{equation}
\lim_{t \rightarrow \infty} \frac{a_{i,j,k}(t)}{a(t)}=\text{constant}>0,
\label{Criterion}
\end{equation} 
where $a(t)=\left[a_i(t)a_j(t)a_k(t) \right]^{\frac{1}{3}}$. An anisotropic spacetime should 
satisfy Eq. (\ref{Criterion}) in order to become isotropic, that is to say, the limit above  
for each scalar factor is necessarily satisfied. 
But for the scale factors (\ref{ai})-(\ref{ajak}), by using Eq. (\ref{Criterion}), that is not the case. For example,
for the $i$-direction, one has
\begin{equation}
\lim_{t \rightarrow \infty} \frac{a_{i}(t)}{a(t)}= \lim_{t \rightarrow \infty} t^{-\frac{1}{3}+\ell},
\end{equation}
and there is no real value for the Lorentz-violation parameter that assures a constant and positive value for 
the isotropization condition---as expected. 

However, as mentioned in Ref. \cite{Gravitation},
an anisotropic model like the Kasner model could turn into an isotropic model  
when the matter content is added.  At a certain
$t$, the matter content dominates the expansion, and the model turns into an isotropic universe as the anisotropies
dilute. In this regard, an isotropic universe could emerge from a highly anisotropic spacetime as
matter is brought to the scene for a mechanism like the inflationary mechanism. 
And the Lorentz symmetry violation would be appropriate before this isotropic period.

\section{Final comments}
Kasner's cosmology is a vacuum solution of the Einstein field equations which describes
an anisotropically expanding universe. In this article, the Kasner geometry in the bumblebee
scenario was studied. Bumblebee gravity concerns a Lorentz symmetry breaking model, in which 
the Lorentz violation arises from the nonzero VEV of the bumblebee field.

Here the bumblebee field is a vector field, a timelike vector field. 
In this scenario, the parameters or exponents of the Kasner metric ($p_1,p_2,p_3$) find out a justification, 
because such parameters present the Lorentz-violating parameter in this scenario of the modified
field equations by the bumblebee field. 
Thus, the anisotropic feature of the Kasner cosmology could have origin in the 
Lorentz symmetry breaking. Moreover, it is argued that our isotropic universe could emerge
from an anisotropic spacetime like the one studied here, where the Lorentz violation is cause of the anisotropies.

\section*{Acknowledgments}
I thank Roberto Maluf for fruitful discussions on Lorentz symmetry breaking.

\end{document}